\documentclass[useAMS,usenatbib,usegraphicx]{mn2e}
\usepackage[T1]{fontenc}
\usepackage{tikz}
\usepackage{aecompl}
\usepackage{color}
\usepackage{epsfig}

\usepackage{gensymb}

\title[A double-peaked H$\alpha$ component in the nucleus of NGC 4958]
  {Detection of a double-peaked H$\alpha$ component from the accretion disc of NGC 4958}
\author[Ricci et al.]
  {T.V.~Ricci\thanks{tiago.ricci@uffs.edu.br}$^{1}$,
  J.E.~Steiner$^2$ \\
  $^1$Universidade Federal da Fronteira Sul, Campus Cerro Largo, RS 97900-000, Brasil \\
  $^2$Instituto de Astronomia, Geof\'isica e Ci\^encias Atmosf\'ericas, Universidade de S\~ao Paulo\\
   }
\date{Released 2002 Xxxxx XX}
\pagerange{\pageref{firstpage}--\pageref{lastpage}} \pubyear{2017}

\def\LaTeX{L\kern-.36em\raise.3ex\hbox{a}\kern-.15em
    T\kern-.1667em\lower.7ex\hbox{E}\kern-.125emX}

\begin{document}

\label{firstpage}

\maketitle

\begin{abstract}

Active Galactic Nuclei are objects associated with the presence of an accretion disc around supermassive black holes found in the very central region of galaxies with a well-defined bulge. In the optical range of the spectrum, a possible signature of the accretion disc is the presence of a broad double-peaked component that is mostly seen in H$\alpha$. In this paper, we report the detection of a double-peaked feature in the H$\alpha$ line in the nucleus of the galaxy NGC 4958. The narrow line region of this object has an emission that is typical of a LINER galaxy, which is the usual classification for double-peaked emitters. A central broad component, related to the broad line region (BLR) of this object, is seen in H$\alpha$ and also in H$\beta$. We concluded that the double-peaked emission is emitted by a circular relativistic Keplerian disc with an inner radius $\xi_{\rm i}$ = 570$\pm$ 83, an outer radius $\xi_{\rm o}$ = 860$\pm$170 (both in units of $GM_{\rm SMBH}/c^2$), an inclination to the line of sight $i$ = 27.2$\degree \pm$0.7$\degree$ and a local broadening parameter $\sigma$ = 1310$\pm$70 km s$^{-1}$. 

\end{abstract}

\begin{keywords}

galaxies: individual: NGC 4958 $-$ galaxies: active $-$ galaxies: nuclei $-$ accretion, accretion discs $-$  techniques: imaging spectroscopy $-$ line: profiles.

\end{keywords}

\section{Introduction} \label{sec:intro}

It has been established that all galaxies with a well-defined bulge contain a supermassive black hole (SMBH) in their centres \citep{2013ARA&A..51..511K}. When matter falls onto the SMBH in the form of an accretion disc, the nucleus of the galaxy becomes very bright and its spectrum reveals the presence of permitted and forbidden lines. This is the so-called Active Galactic Nucleus (AGN). Quasars are among the most luminous types of AGNs in the Universe, while Seyferts and LINERs belong to the low luminosity branch of the AGN population \citep{2008ARA&A..46..475H}. 

The optical spectra of AGNs are usually separated into two types: those with broad permitted and narrow forbidden emission lines (type 1) and those containing only narrow lines (type 2). Moreover, some galaxies also show double-peaked profiles, mostly seen in H$\alpha$, which have been related to the emission of the accretion disc around the SMBHs. One of the first galaxies where this feature was detected is Arp 102B \citep{1989ApJ...344..115C, 1989ApJ...339..742C}. Since then, other objects have revealed this component, like NGC 1097 \citep{1993ApJ...410L..11S}, NGC 7213 \citep{2014MNRAS.438.3322S}, NGC 4450 \citep{2000ApJ...541..120H} and NGC 4203 \citep{2000ApJ...534L..27S}. An analysis of a sample of double-peaked emitters, selected among radio-loud galaxies, has been made by \citet{1994ApJS...90....1E, 2003ApJ...599..886E}. They concluded that relativistic disc models are the best explanation for the double-peaked feature, ruling out other explanations as broad line regions (BLRs) around a binary black hole system, outflows or spherically symmetric BLR illuminated by an anisotropic ionizing source. They also claim that double-peaked emitters are mostly classified as LINERs. \citet{2003AJ....126.1720S} detected 116 double-peaked emitters in a sample of 3212 AGN galaxies selected from the Sloan Digital Sky Survey (SDSS). They also proposed that all 116 sources contain relativistic disc emission and that the [O I]/[O III] ratios are usually larger in double-peaked emitters when compared to their parent AGN sample; this is a typical characteristic of a LINER-like spectrum. Although most of the 116 sources are radio quiet ($\sim$ 76 \%), these authors noticed that it is 1.6 times more likely that the double-peaked emitters are radio sources.  

\begin{figure*}
\includegraphics[scale=0.9]{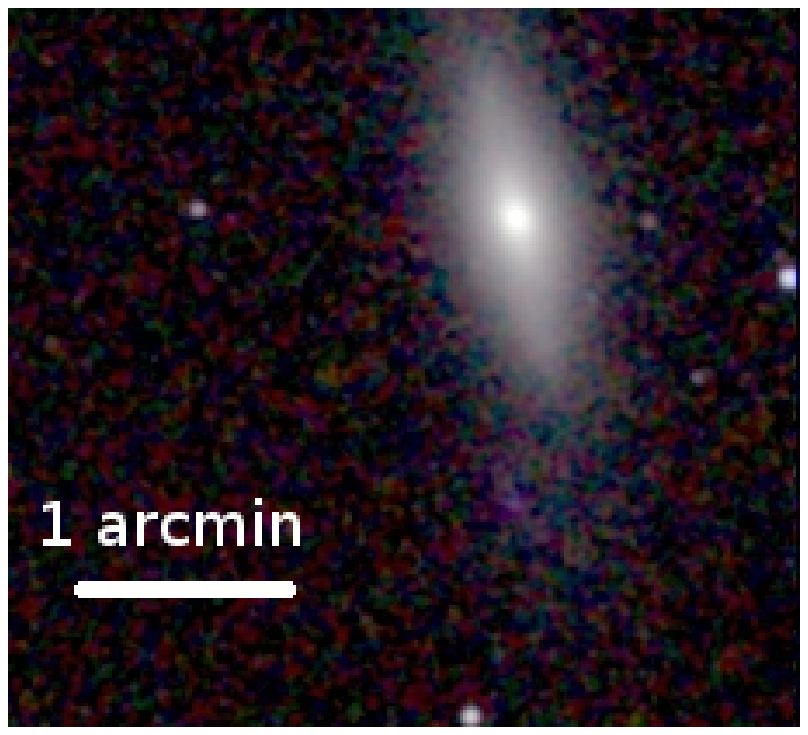} 
\includegraphics[scale=0.5]{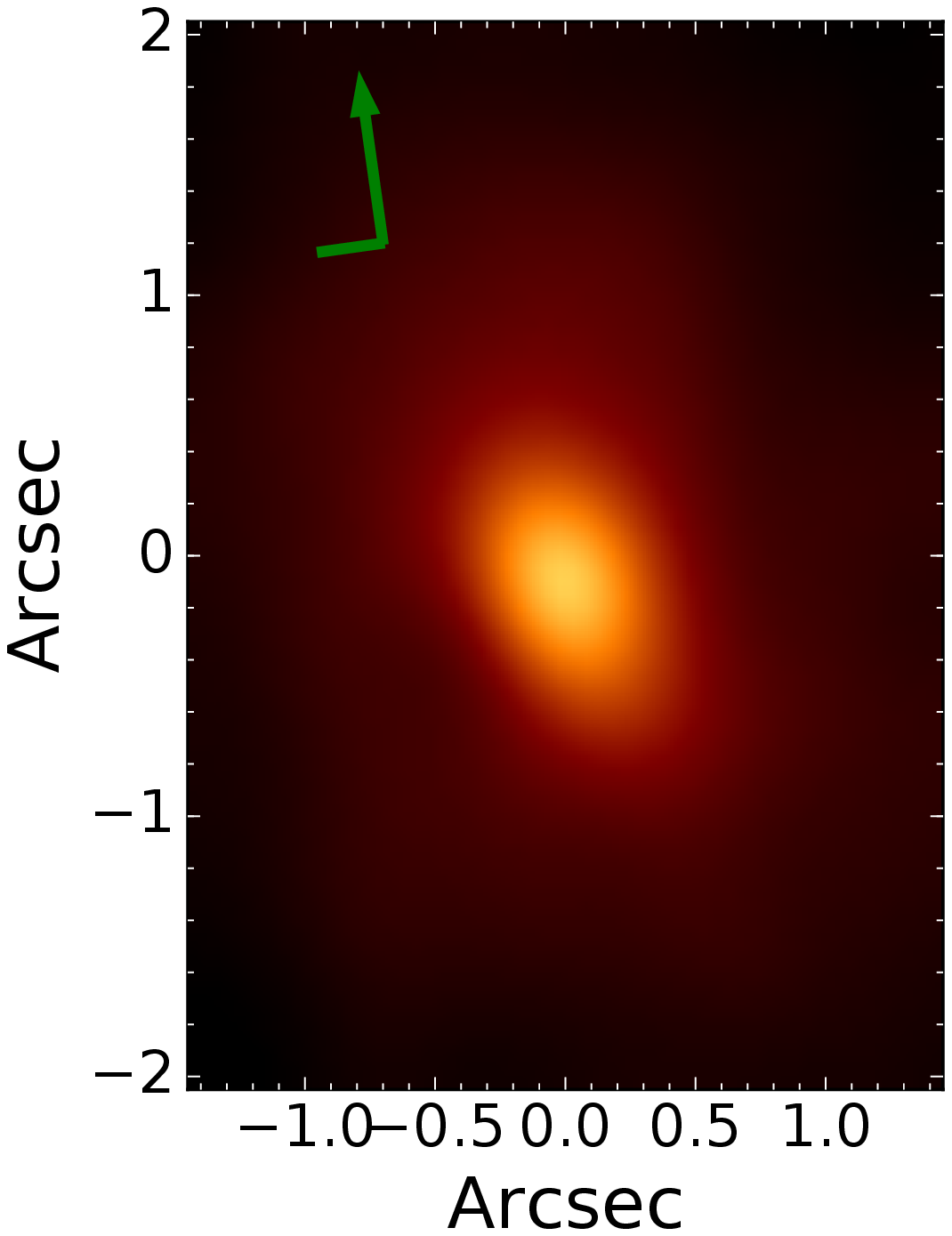}
\caption{Left: composed JHK image of NGC 4958 from the Two Micron All Sky Survey (2MASS) \citep{2006AJ....131.1163S}. Right: image of the stellar continuum ($\sim$ 5500 \AA) of the GMOS data cube of NGC 4958 analysed in this paper. For this galaxy, 1 arcsec = 64 pc. \label{fig:ngc4958}}
\end{figure*}

The main goal of this paper is to report the detection of a double-peaked H$\alpha$ component in the galaxy NGC 4958 using high spatial resolution observations made with the Integral Field Unit (IFU) of the Gemini Multi-Object Spectrograph (GMOS). NGC 4958 is an edge-on SB0 galaxy (see Fig. \ref{fig:ngc4958}) with a magnitude $m_{\rm B}$ = 11.6 \citep{1991trcb.book.....D}, located at a distance of 13.2 Mpc \citep{2007A&A...465...71T}. \citet{1989A&A...226...23B} detected weak nuclear emission lines in this galaxy (mainly H$\alpha$, [N II] and [S II]). A compact radio source is seen in the centre of NGC 4958 \citep{1998AJ....115.1693C}, with a power of 1.9$\times$10$^{20}$ W/Hz at 1.4 GHz \citep{2011ApJ...731L..41B}. Both the radio and optical emission are in accordance with a non-thermal emission from a LINER-like AGN \citep{2008ARA&A..46..475H}. The fact that NGC 4958 is a low luminosity nearby galaxy makes it an interesting case to study its double-peaked feature.

The paper is organized as it follows: Section \ref{sec:obs_data_reduction} describes the observations and data reduction. Section \ref{sec:spectral_synthesis} presents the spectral synthesis applied to the data cube of NGC 4958 in order to subtract the stellar component of the galaxy. In Section \ref{sec:narrow_line_profiles}, we characterize the emission from the narrow line region (NLR), the BLR and also from the double-peaked H$\alpha$ component. Section \ref{sec:modelling_double_peaked_profile} is related to the final model of the double-peaked H$\alpha$ profile. Section \ref{sec:discussion} presents a discussion about the results obtained for the nuclear region of NGC 4958. Section \ref{sec:conclusions} shows our main conclusions. 

\section{Observations and data reduction} \label{sec:obs_data_reduction}

NGC 4958 was observed on 2015 May 12 with the Gemini South Telescope (programme GS-2015A-Q-35), using the GMOS-IFU instrument \citep{2002PASP..114..892A, 2004PASP..116..425H} in the one-slit mode. In this setup, 500 microlenses with a diameter of 0.2 arcsec each are placed at the focal plane of the telescope, covering a field of view (FOV) of 3.5 $\times$ 5.0 arcsec$^2$. The sky is observed simultaneously using 250 microlenses located 1 arcmin away from the science FOV. The microlenses slice the image of the target and direct the light to a set of fibres that are arranged linearly at the nominal position of the spectrograph slit (pseudo-slit). Finally, the spectrum of each spatial position of the FOV is recorded in a mosaic of three CCDs. Three 582 s exposures were taken for this object. The B600-G5323 grating was used, with a central wavelength at 5620 \AA. This covered a spectral range of 4020 to 7200 \AA. The spectral resolution is 1.8 \AA (full width at half-maximum, FWHM), as estimated from the CuAr lamp and from the skylines. The seeing of the observations was estimated to be 0.45 arcsec using stars that are present in the acquisition image obtained with the GMOS imager in the r filter (SDSS system).

Each exposure was reduced with the standard Gemini {\sc iraf} packages. Bias, flat-fields, wavelength calibration, dispersion correction and flux calibration procedures were applied to the science data. Cosmic rays were removed using the {\sc lacos} algorithm \citep{2001PASP..113.1420V}. We then built three data cubes, one for each exposure, which were corrected for the differential atmospheric refraction effect using an algorithm developed by us and based on the equations by \citet{1982PASP...94..715F} and \citet{1998Metro..35..133B}. After that, we averaged all three data cubes. 

In addition to the basic reduction procedures, we applied complementary techniques in order to improve the quality of the data cube. First, we removed high-frequency noises from the spatial dimension using a Butterworth filter with a cut-off frequency of 0.30 $F_{\rm NY}$, where $F_{\rm NY}$ is the Nyquist frequency, and a filter order n = 2. Then, low-frequency instrumental fingerprints were removed using the PCA Tomography \citep{2009MNRAS.395...64S}. At this stage, the telluric lines were removed and after that, the data cube was corrected for the Milky Way extinction using the curves by \citet{1989ApJ...345..245C} and assuming $A_{\rm V}$ = 0.129 \citep{2011ApJ...737..103S}. Finally, we deconvolved the spatial dimension of the data cube using the Richardson Lucy technique \citep{1972JOSA...62...55R, 1974AJ.....79..745L} using a Moffat Point Spread Function (PSF) with a FWHM = 0.45 arcsec and an index $\beta$ = 2.9. The resulting PSF, after 10 iterations with the Richardson Lucy procedure, has a FWHM = 0.41 arcsec (26 pc), as estimated from an image extracted from the red wing of H$\alpha$. An image extracted from the stellar continuum ($\sim$ 5500 \AA) of the final data cube of NGC 4958 is shown in Fig. \ref{fig:ngc4958}. A detailed description of these techniques applied to GMOS-IFU data is presented in \citet{2019MNRAS.483.3700M, 2014MNRAS.438.2597M,2015MNRAS.450..369M}.

\section{Spectral synthesis} \label{sec:spectral_synthesis}

\begin{figure}
\centering \includegraphics[scale=0.5]{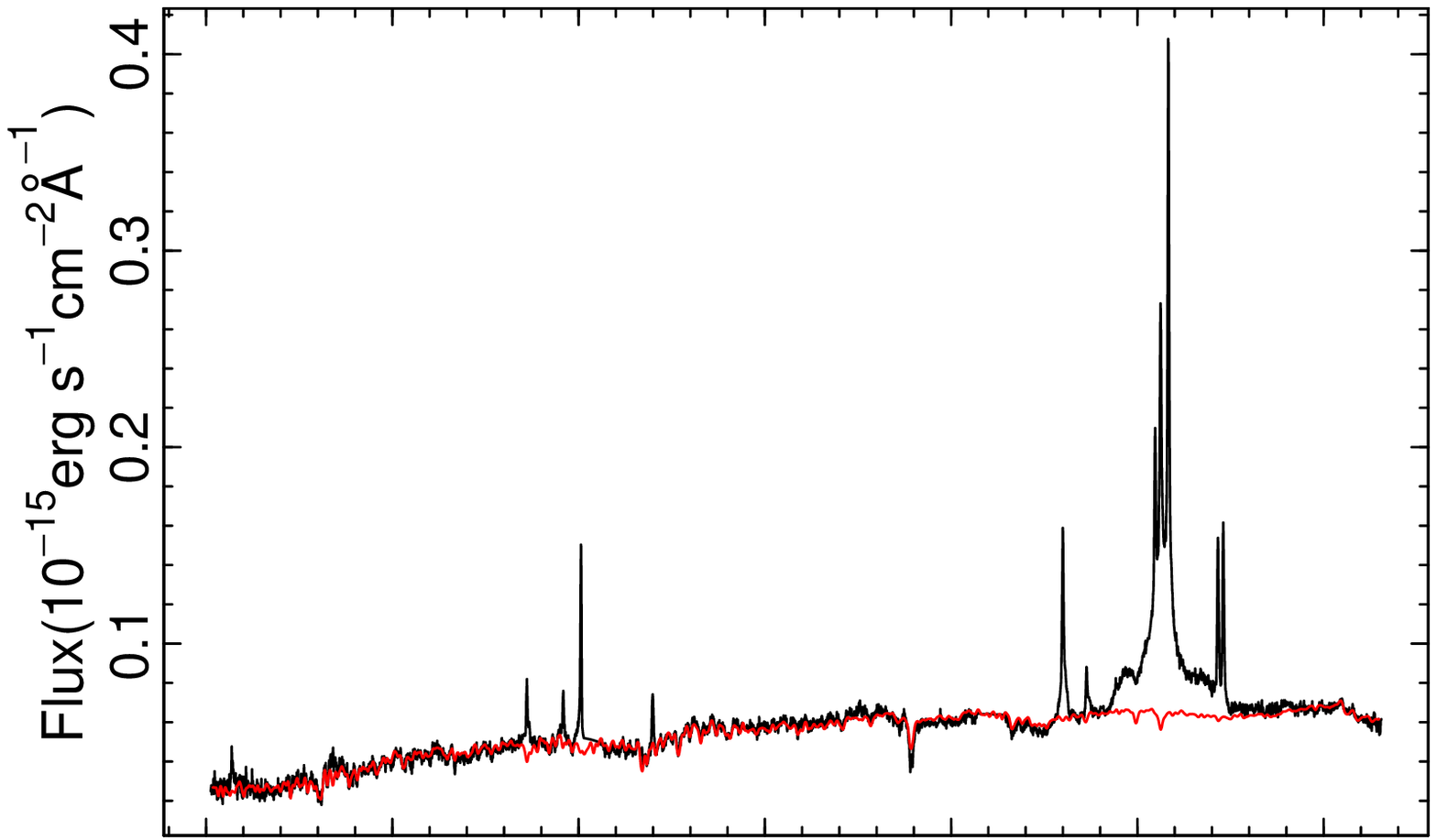} 
\centering \includegraphics[scale=0.5]{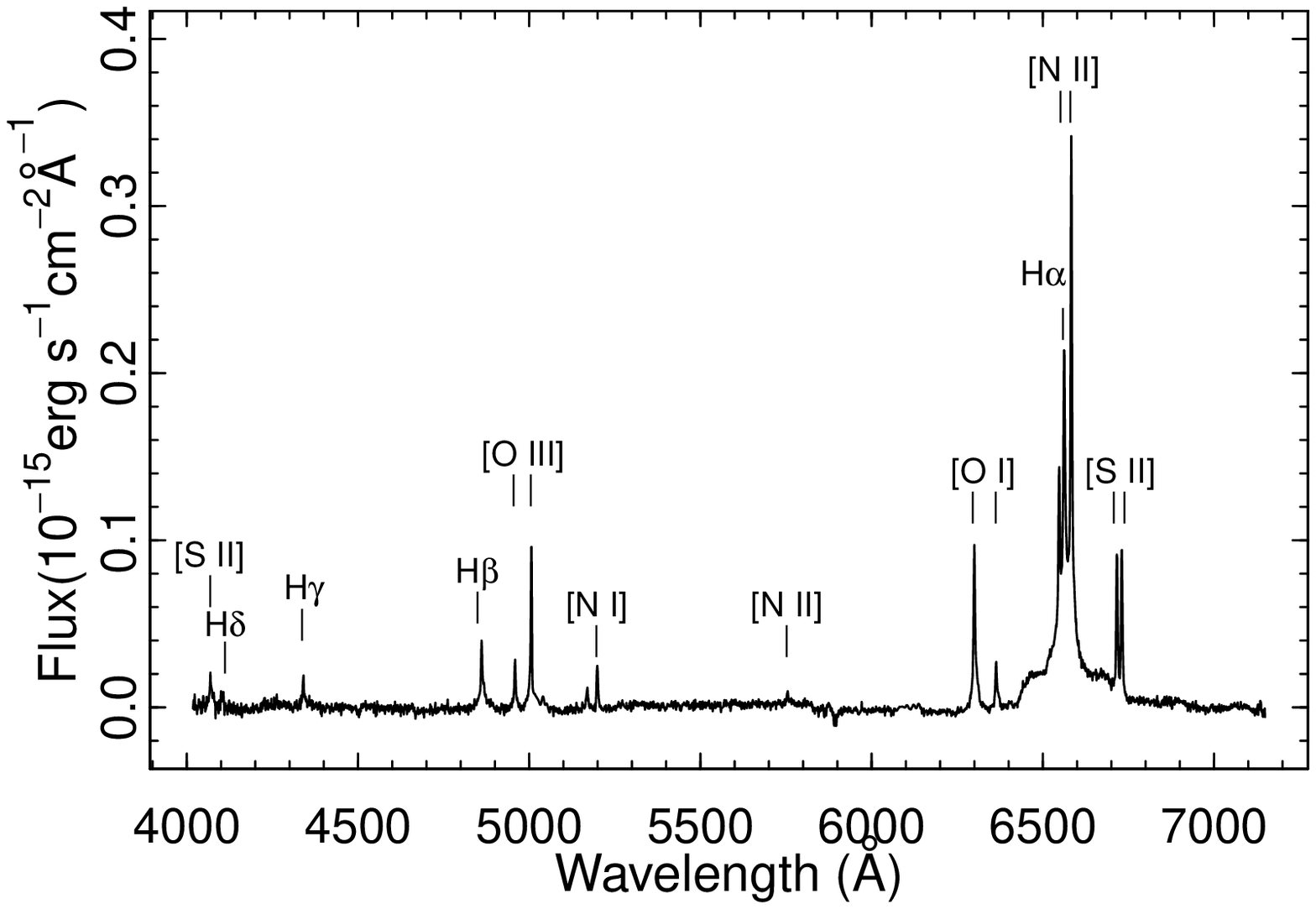} 
\caption{Top: observed spectrum of a nuclear spaxel of the GMOS data cube of NGC 4958 (black) together with the spectral synthesis result (red), obtained with the {\sc starlight} software. Botton: residual spectrum revealing all nuclear emission lines of NGC 4958. A broad H$\alpha$ component is clearly seen in the spectrum even before the starlight subtraction. \label{fig:nuc_spectrum}}
\end{figure}

In order to analyse the nuclear gas emission of NGC 4958, we subtracted the stellar component from each spaxel of the data cube of this object by means of a spectral synthesis. This procedure was performed with the {\sc starlight} software \citep{2005MNRAS.358..363C}. The simple stellar populations (SSP) used here are described in \citet{2015MNRAS.449.1177V}. Since NGC 4958 is an early-type galaxy, all SSPs in the basis are older than 30 Myr and have metallicities between 0.2 and 2.4 $Z_\odot$, an $\alpha_{\rm enh}$ = 0.0 and 0.4, and an initial mass function of \citet{2003PASP..115..763C}. We also used the \citet{1989ApJ...345..245C} curve to take the internal extinction of this object into account. In Fig. \ref{fig:nuc_spectrum}, we show the spectrum that corresponds to the spaxel located at the centre of the galaxy with the spectral synthesis result superposed on it, and also the residual spectrum containing only the nuclear gas emission. In Appendix \ref{sec:spectral_synthesis_results}, we briefly discuss the characteristics of the SSPs that best fitted the spectra of the data cube of NGC 4958.

We extracted the nuclear spectrum from the data cube of NGC 4958 containing only emission lines by summing the four spectra located in the position of the stellar photometric centre of this object and correcting the flux by assuming a point-like source with a PSF = 0.41 arcsec. This procedure avoids possible contamination from circumnuclear emissions. 

\section{Disetangling emission line components} \label{sec:narrow_line_profiles}

The presence of a broad H$\alpha$ component in the nucleus of NGC 4958 is clear even before the starlight subtraction (see Fig. \ref{fig:nuc_spectrum}). This feature ranges from $\sim$ 6380 to 6750 \AA. However, to avoid the emission of the narrow component of the H$\alpha$+[N II]$\lambda \lambda$6548, 6583 lines, a first attempt to fit a relativistic keplerian disc model was made only between 6380 and 6500 \AA\ and between 6640 and 6750 \AA. Moreover, the [S II]$\lambda \lambda$6714, 6731 doublet was removed with a simple linear interpolation between two convenient wavelengths. We finally added Gaussian random noise to the interpolation results. 

\begin{figure}
\centering \includegraphics[scale=0.5]{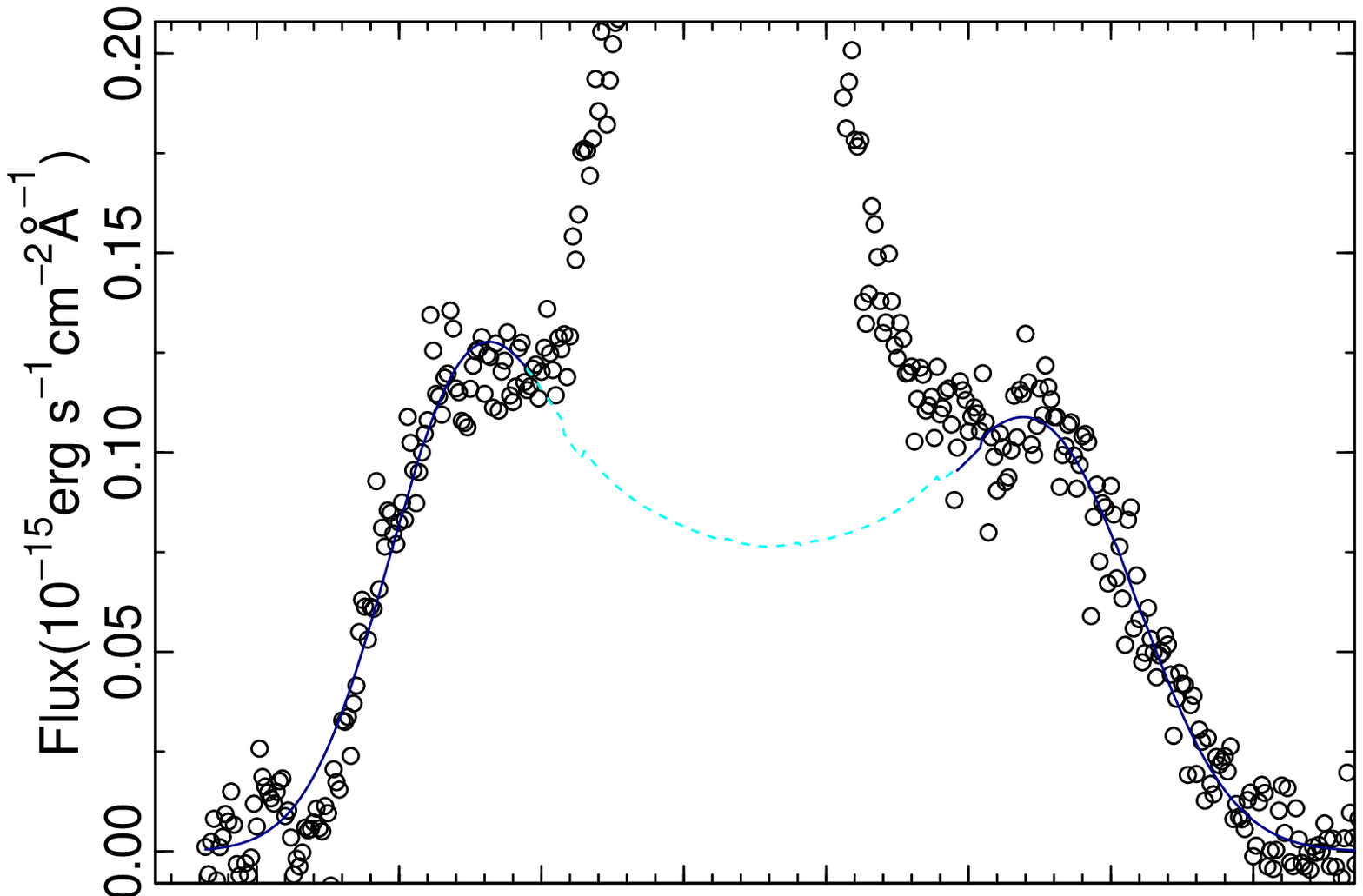} 
\centering \includegraphics[scale=0.5]{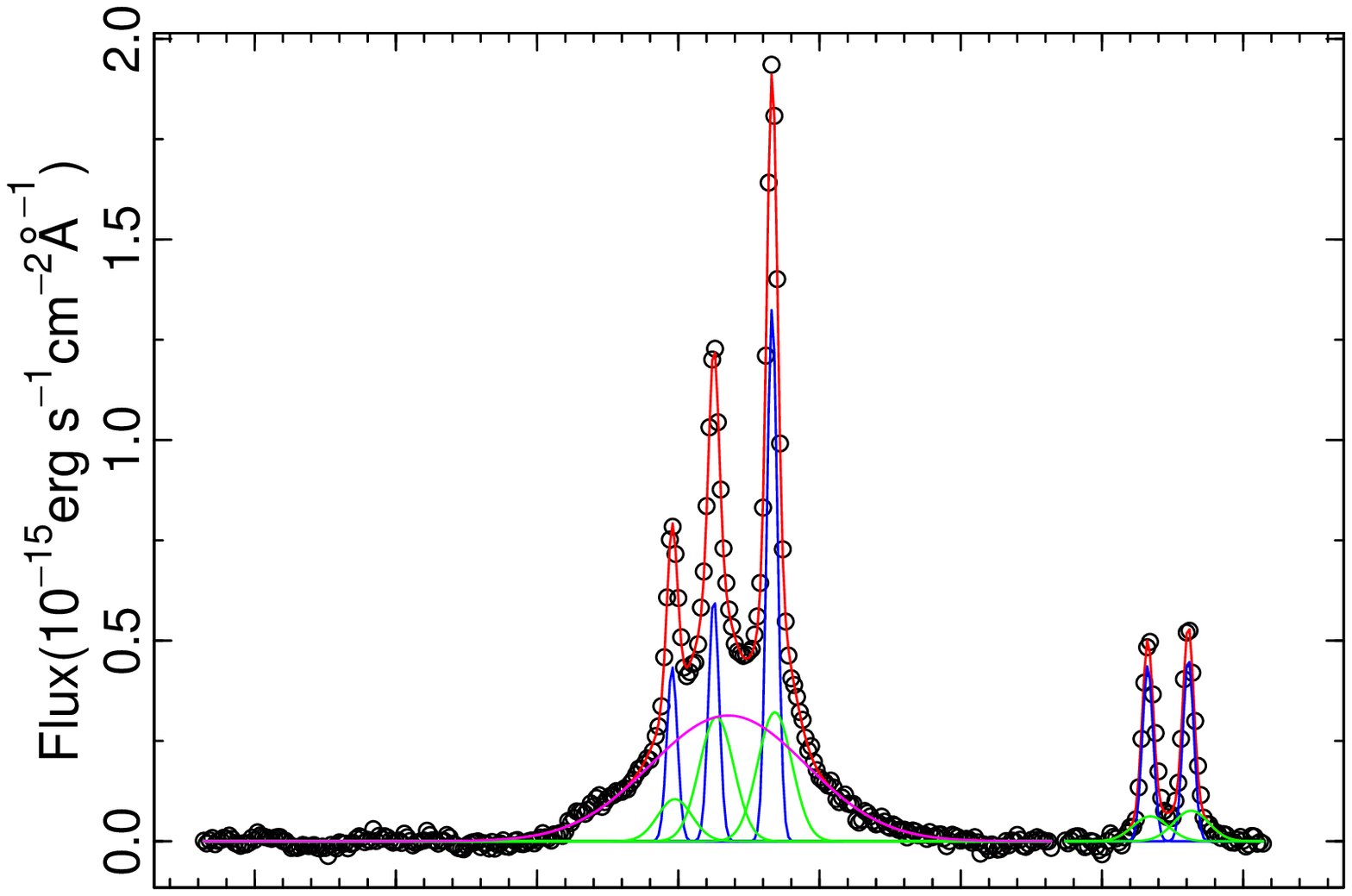} 
\centering \includegraphics[scale=0.5]{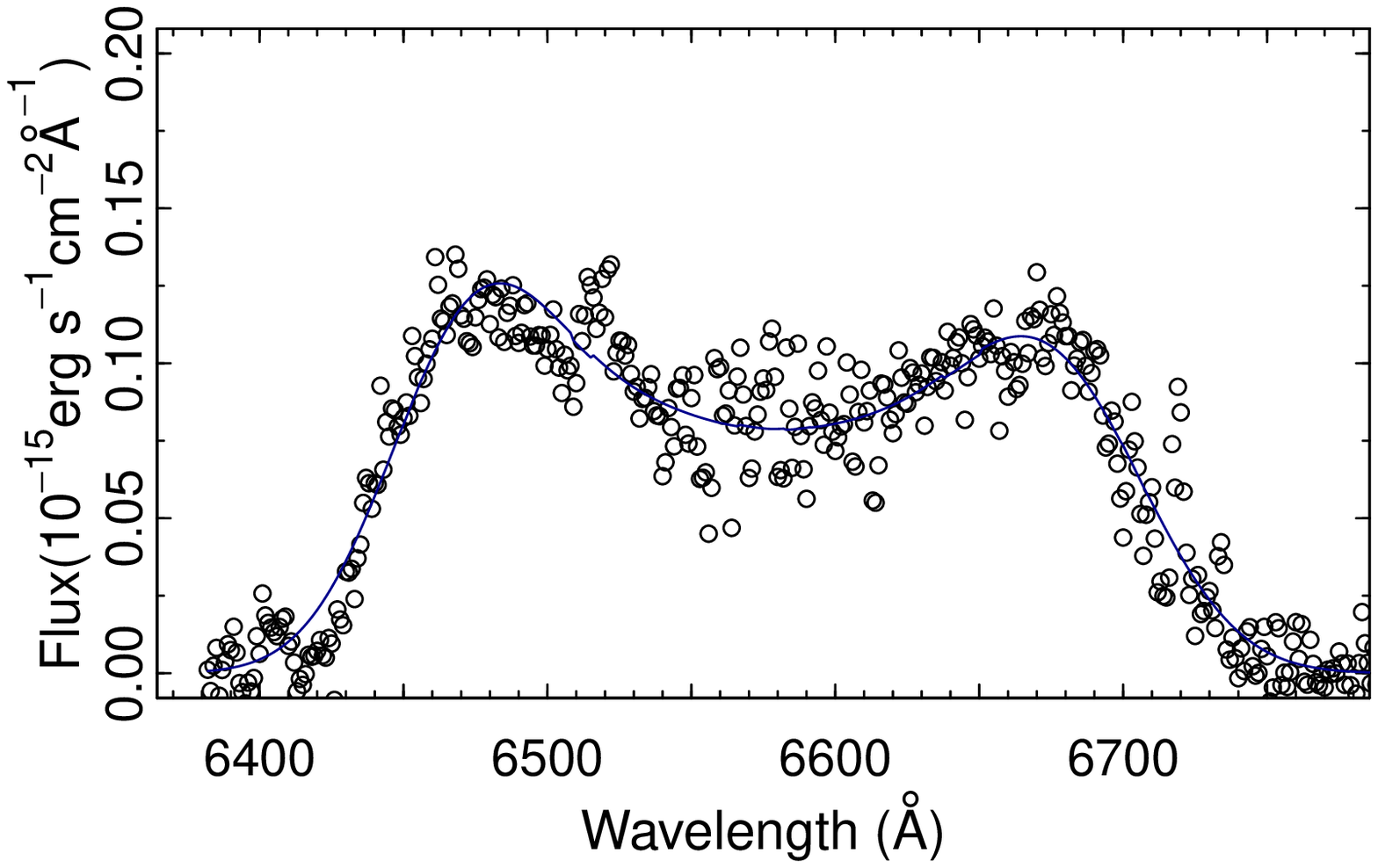} 

\caption{Top: first attempt to fit a circular relativistic keplerian model in the H$\alpha$ line of the nucleus of NGC 4958. The dark blue full line represents the double-peaked profile in the spectral region where the fit was performed (see text for details), while the cyan dashed line is the same model result, but in the spectral region where a zero weight was given to the fitting procedure. Middle: Fitted profiles of the narrow components of the [N II]$\lambda \lambda$6548, 6584, H$\alpha$ and [S II]$\lambda \lambda$6716, 6731 lines. Note that there are two sets of Gaussian functions to fit the narrow components of these emissions: Set 1 is shown in blue and Set 2 is shown in green. All Gaussian functions from a given set have the same kinematics (radial velocities and FWHM). The central broad component of H$\alpha$ is shown in magenta. Bottom: circular relativistic keplerian model fitted to the H$\alpha$ line. For this case, we subtracted the contribution of the narrow and central broad components of the [N II]$\lambda \lambda$6548, 6584, H$\alpha$ and [S II]$\lambda \lambda$6716, 6731 lines before fitting the double-peaked profile. The dark blue full line is related to the model result in the spectral region where the fit was performed. \label{fig:disc_steps}}
\end{figure}

We used the relativistic disc model proposed by \citet{1989ApJ...344..115C} and \citet{1989ApJ...339..742C}. This model assumes that an ionizing source is located at the centre of a circular disc and that the line emission takes place in a thin ionized layer at the surface of this disc. The calculations take into account some relativistic phenomena, as the gravitational redshift, the transversal redshift and light bending by a weak gravitational field. We assumed an emissivity law for H$\alpha$ given by $\epsilon$($\xi$) $\propto$ $\xi^{-q}$, where $\xi \equiv R/R_{\rm grav}$ with $R$ being the radius of the disc and $R_{\rm grav} \equiv GM_{\rm SMBH}/c^2$ is the gravitational radius, $G$ is the gravitational constant, $M_{\rm SMBH}$ is the mass of the supermassive black hole and $c$ is the speed of light. We fixed $q$ = 3, since this value is supported by photoionization calculations applied to accretion discs \citep{1989A&A...213...29C, 1990A&A...229..292C, 1990A&A...229..302D, 1990A&A...229..313D}. We fitted the relativistic disc model with the Levenberg-Marquardt algorithm, using as free parameters the inner and outer radii ($\xi_i$ and $\xi_o$, respectively), the inclination $i$ of the disc to the line of sight, the broadening parameter $\sigma$ (to account for the velocity dispersion in a given region of the disc, probably related to turbulence), and a normalization factor to take the flux of the line into account. The resulting profile of the broad H$\alpha$ line is shown in the top panel of Fig. \ref{fig:disc_steps}. It is clear that this profile reveals a double-peaked component typical of relativistic discs (see e.g. \citealt{1989ApJ...344..115C, 1989ApJ...339..742C, 1994ApJS...90....1E,2003ApJ...599..886E, 2003AJ....126.1720S, 2017ApJ...835..236S}).

To characterize the emission from the NLR, we first subtracted the double-peaked H$\alpha$ component fitted with the preliminary relativistic disc model. The resulting spectrum is seen in the middle panel of Fig. \ref{fig:disc_steps}. Note that, in addition to the narrow components, one can also see an apparent broad H$\alpha$ component, which is possibly associated with a BLR. From now on, we will call this feature as the central broad component to avoid confusion with the double-peaked component. We then fitted the line profiles as a sum of Gaussian functions using the Levenberg-Marquardt algorithm. We started the fitting procedure with the [N II]+H$\alpha$ emission, using two sets of Gaussian functions to adjust the narrow components (i.e. two Gaussian functions for each line) and one Gaussian function to take the central broad component into account. We assumed that each set of narrow Gaussian functions has only one radial velocity and FWHM for all lines, i.e. there are only two free kinematic parameters per set. In this case, Set 1 corresponds to the Gaussian functions with lower FWHM values while Set 2 is related to the Gaussian functions with higher FWHM results. The amplitudes of all narrow Gaussian functions were taken as free parameters, but fixing the [N II]$\lambda$6583/[N II]$\lambda$6548 ratio to the theoretical value of 3.06 \citep{2006agna.book.....O}. For the central broad component, the amplitude, radial velocity and FWHM of its corresponding Gaussian function were set as free parameters. The [S II] $\lambda \lambda$6714, 6731 doublet was fitted using the same kinematics found for the [N II]+H$\alpha$ lines, i.e., the free parameters here were just the amplitudes associated with each set of Gaussian functions for both [S II] lines. The fitted profiles are also shown in the middle panel of Fig. \ref{fig:disc_steps}. 
 
\begin{figure*}
\centering \includegraphics[scale=0.5]{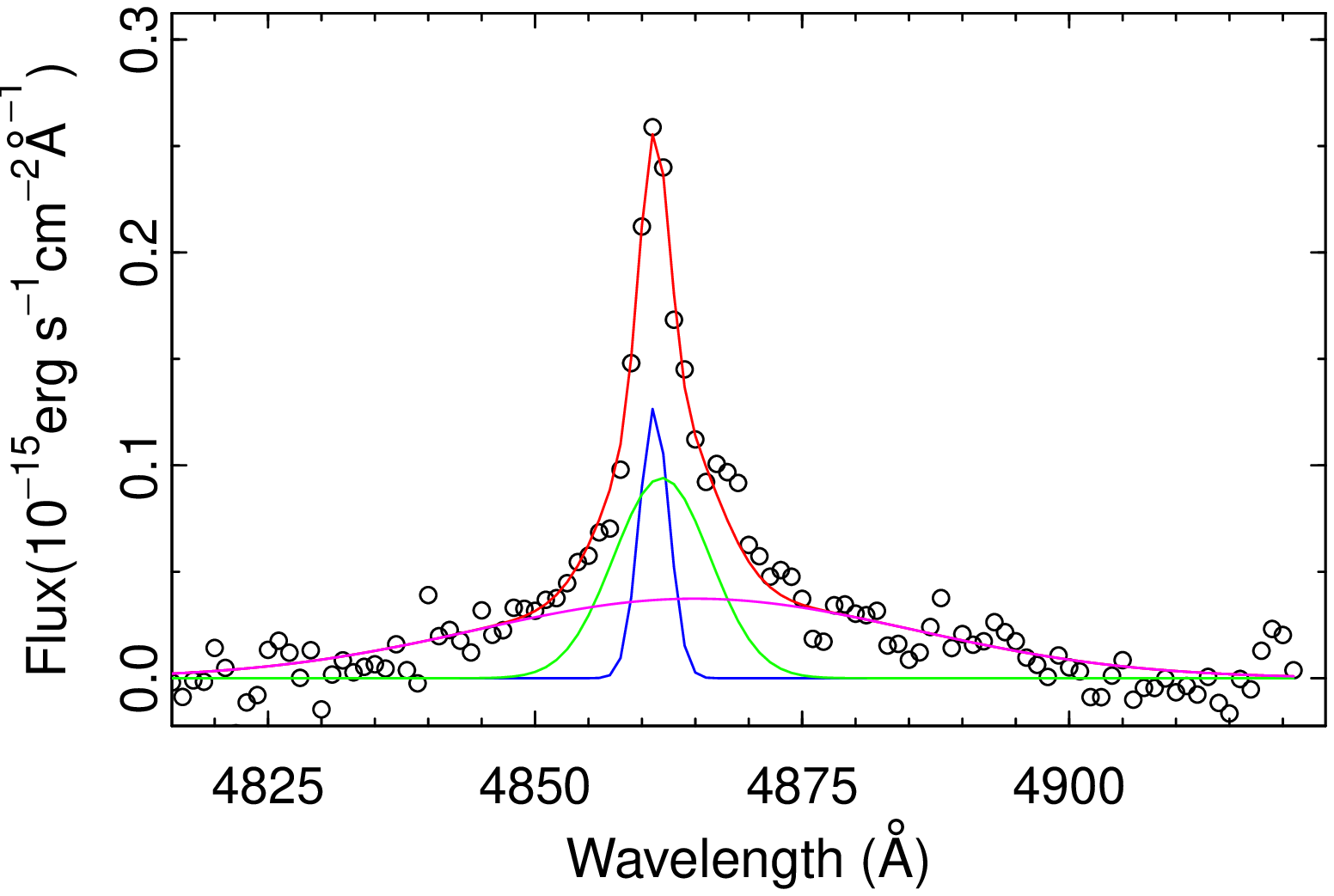} 
\centering \hspace{-0.3cm}\includegraphics[scale=0.5]{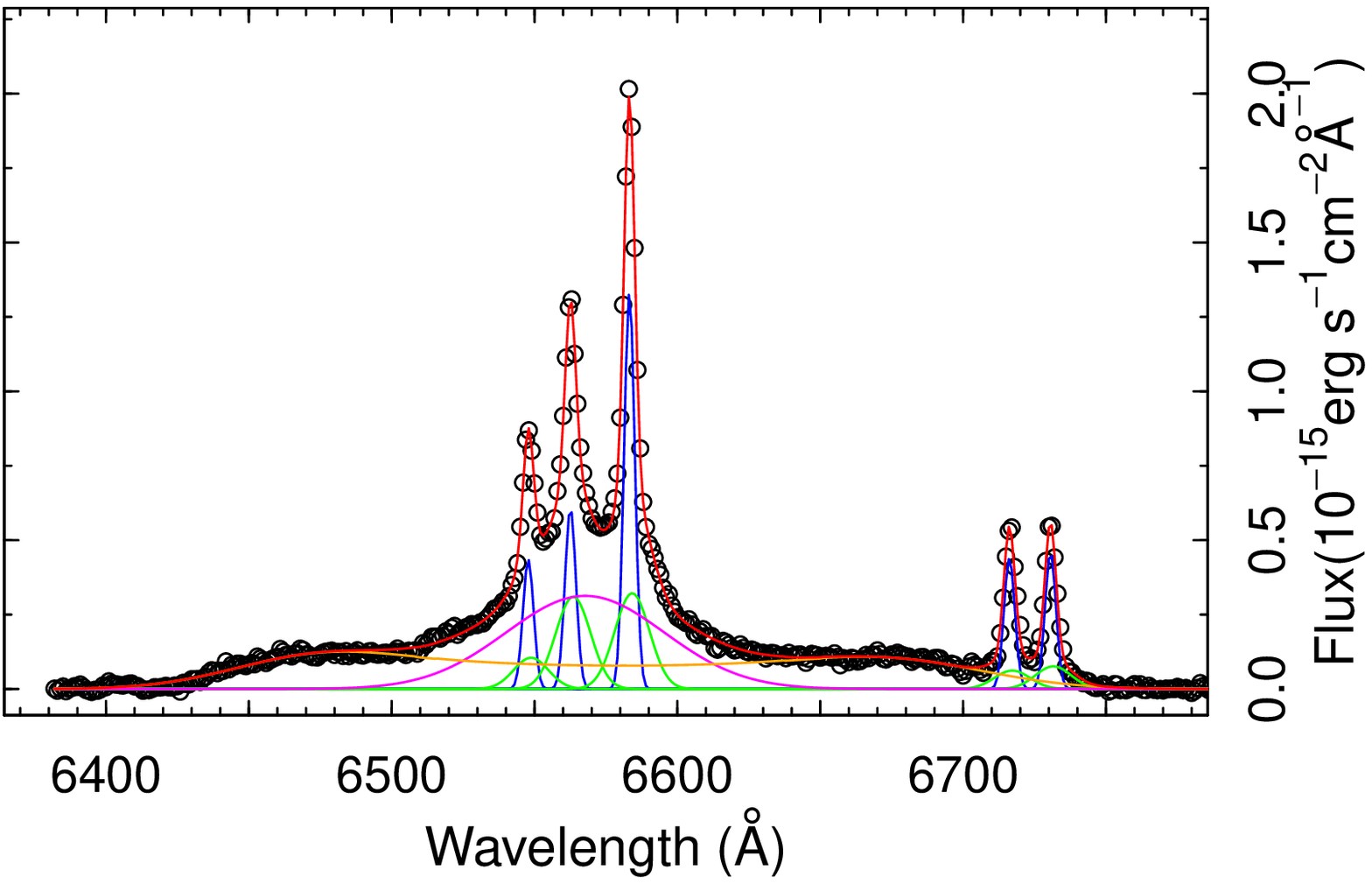} 
\caption{Left: Observed H$\beta$ emission of the nucleus of NGC 4958. The profile of this line was fitted using two sets of Gaussian lines to account for the narrow component of the line and one Gaussian function for the central broad component. The kinematics of all Gaussian functions fitted here were fixed to the same values found for the H$\alpha$ line. Right: Red portion of the nuclear spectrum, showing the [N II]$\lambda \lambda$6548, 6583, H$\alpha$ and [S II]$\lambda \lambda$6716, 6731 line emissions. The final profile of these lines is composed by the sum of the emissions from the NLR, the BLR and the accretion disc. \label{fig:nuc_spectrum_lines}}
\end{figure*}

The H$\beta$ line profile was fitted using also two sets of narrow Gaussian functions plus a broad Gaussian function for the central broad component (no double-peaked feature is apparent in this line). The free parameters here were the amplitude of each Gaussian function since we assumed that H$\beta$ has the same kinematics as H$\alpha$. The fitted profiles are shown in Fig. \ref{fig:nuc_spectrum_lines}. 
 
In order to classify the nuclear emission of NGC 4958, we calculated the fluxes of the narrow components of the H$\alpha$, H$\beta$, [N II]$\lambda$6584 and [S II]$\lambda \lambda$6716, 6731 lines using the Gaussian function profile results. For the other emission lines seen in the spectrum, as [O I]$\lambda$6300, [O III]$\lambda$5007 and [N II]$\lambda$5755, the fluxes were calculated with direct numerical integration between two convenient wavelengths. Typical line ratios, the observed H$\alpha$ flux F(H$\alpha$), and the kinematic information (radial velocity V$_r$ and FWHM) found for both sets of narrow Gaussian functions are shown in Table \ref{tab:nlr}. These results for the NLR of NGC 4958 suggest that this galaxy has a LINER-like AGN \citep{1980A&A....87..152H,1983ApJ...269L..37H,1983ApJ...264..105F,1997ApJS..112..315H,2006MNRAS.372..961K}.


\begin{table}
\caption{NLR parameters for NGC 4958. The radial velocities V$_{\rm Rn}$ are shown with respect to the heliocentric velocity V$_{\rm HEL}$ = 1455 km s$^{-1}$ presented in \citet{2003AJ....126.2268W}. All errors presented for the parameters below are purely statistical. \label{tab:nlr}}

\begin{tabular}{lc}
\hline
Parameter & Observed value \\
\hline
\ F(H$\alpha$) (10$^{-15}$ erg s$^{-1}$ cm$^{-2}$) & 7.3$\pm$0.3 \\
\ H$\alpha$/H$\beta$ & 5.0$\pm$0.4 \\
\ E(B-V) & 0.46$\pm$0.07 \\
\ Log L(H$\alpha$) (erg s$^{-1}$) & 38.7$\pm$0.5 \\
\ Log L$_{\rm bol}$ (erg s$^{-1}$) & 42.4 \\
\ Log L$_{\rm bol}$/L$_{\rm edd}$ & -4.3 \\
\ [N II]$\lambda6584$/H$\alpha$ & 1.49$\pm$0.08 \\
\ [S II]/H$\alpha$&0.88$\pm$0.04 \\
\ [O I]$\lambda6300$/H$\alpha$&0.77$\pm$0.03 \\
\ [O III]$\lambda5007$/H$\beta$ &2.7$\pm$0.2 \\
\ V$_{\rm R1}$ (km s$^{-1}$) &-9$\pm$1 \\
\ FWHM$_{\rm 1}$ (km s$^{-1}$)&175$\pm$3 \\
\ V$_{\rm R2}$ (km s$^{-1}$)&32$\pm$7 \\
\ FWHM$_{\rm 2}$ (km s$^{-1}$)&627$\pm$24 \\
\ T$_{\rm e}$ [N II] (K) & 1.6$\times$10$^4$\\
\ n$_{\rm e}$ [S II] (cm$^{-3}$) & 660$\pm$140\\
\hline 
\end{tabular}

\end{table}

In addition to line fluxes, we also show in Table \ref{tab:nlr} the colour excess E(B-V) calculated with the observed H$\alpha$/H$\beta$ ratio, assuming an intrinsic value of 3.1 \citep{1983ApJ...269L..37H,1983ApJ...264..105F}. The H$\alpha$ luminosity  L(H$\alpha$) was calculated using the derredened H$\alpha$ flux. The bolometric luminosity L$_{\rm bol}$ was calculated using the [O I] and [O III] line luminosities according to the bolometric correction presented in \citet{2009MNRAS.399.1907N,2013peag.book.....N}. The Eddington ratio Log L$_{\rm bol}$/L$_{\rm edd}$ was obtained assuming M$_{\rm SMBH}$ = 1$\times$10$^8$ M$_\odot$, calculated with the M$_{\rm SMBH} \times \sigma$ relation \citep{2013ARA&A..51..511K} using a central stellar velocity dispersion of 154 km s$^{-1}$, obtained from Hyperleda \citep{2014A&A...570A..13M}.  The electron density (n$_{\rm e}$) and temperature (T$_{\rm e}$) were calculated with the [S II] and [N II] lines, respectively, using the {\sc temden} code \citep{1995PASP..107..896S}, which is part of the {\sc nebular} package in {\sc iraf}. All values are also in agreement with a photoionized medium that is typical of a LINER-like nucleus.

We also measured a few parameters to characterize the central broad component ({\sc cbc}), as the observed H$\alpha$ flux F(H$\alpha$)$_{\rm cbc}$, the (H$\alpha$/H$\beta$)$_{\rm cbc}$ ratio, the redshift  ($\Delta \lambda$/$\lambda$)$_{\rm cbc}$ and the FWHM$_{\rm cbc}$. The results are shown in Table \ref{tab:blr}. 

\begin{table}
\caption{Parameters related to the central broad component ({\sc cbc}) and to the double-peaked component ({\sc dpc}). Both ($\Delta \lambda$/$\lambda$) measurements are shown with respect to the rest-frame reference ($\lambda_0$ = 6562.8 \AA). All errors presented for the parameters below are purely statistical. \label{tab:blr}}

\begin{tabular}{lc}
\hline
Parameter & Observed value \\
\hline
\ F(H$\alpha$)$_{\rm cbc}$ (10$^{-15}$ erg s$^{-1}$ cm$^{-2}$) & 22$\pm$1 \\
\ (H$\alpha$/H$\beta$)$_{\rm cbc}$ & 11$\pm$1\\
\ ($\Delta \lambda$/$\lambda$)$_{\rm cbc}$ &0.00074$\pm$0.00005 \\
\ FWHM$_{\rm cbc}$ (km s$^{-1}$)&2990$\pm$60 \\
\ F$_{\rm red}$(H$\alpha$)$_{\rm dpc}$ (10$^{-15}$ erg s$^{-1}$ cm$^{-2}$) & 13$\pm$1 \\
\ F$_{\rm blue}$(H$\alpha$)$_{\rm dpc}$/F$_{\rm red}$(H$\alpha$)$_{dpc}$ & 1.2$\pm$0.1\\
\ FWHM$_{\rm dpc}$ (km s$^{-1}$)&12000$\pm$340 \\
\ ($\Delta \lambda$/$\lambda$)$_{\rm dpc}$ &0.0016$\pm$0.0005 \\
\ $\lambda_{\rm blue}$ (km s$^{-1}$)& -4100$\pm$210\\
\ $\lambda_{\rm red}$ (km s$^{-1}$)& 4900$\pm$210\\

\hline 
\end{tabular}

\end{table}

After the characterization of the NLR and the BLR, we were able to subtract the emissions of these components from the nuclear spectrum. What is left here is the observed double-peaked component ({\sc dpc}), shown in the bottom panel of Fig. \ref{fig:disc_steps}. In order to characterize this feature, we measured the fluxes of the red and the blue regions (F$_{\rm red}$(H$\alpha$)$_{\rm dpc}$ and F$_{\rm blue}$(H$\alpha$)$_{\rm dpc}$, respectively) by doing a direct numerical integration bluewards and redwards, respectively, from the central $\lambda$ = 6573 \AA\ of this component. This central wavelength also corresponds to the measured redshift [($\Delta \lambda$/$\lambda$)$_{\rm dpc}$] of the double-peaked component. Then, we measured the positions of the peaks of the blue and red components ($\lambda_{\rm blue}$ and $\lambda_{\rm red}$). Finally, the FWHM of the double-peaked component was determined using as the maximum emission the average of the flux densities measured for both the red and the blue peaks. The results are shown in Table \ref{tab:blr}. 

\section{Modelling the double-peaked profile} \label{sec:modelling_double_peaked_profile}

Once we isolated the emission of the double-peaked component in the nuclear spectrum, we are now in the position to model this feature across the whole 6380 - 6750 \AA\ range using again the circular relativistic disc model proposed by \citet{1989ApJ...344..115C} and \citet{1989ApJ...339..742C}. Although more realistic (and also more complex) models are available in the literature [e.g. the elliptical disc model \citep{1995ApJ...438..610E} and the spiral arm model \citep{1999ASPC..175..189G,2012ApJ...748..145S, 2017ApJ...835..236S, 2017MNRAS.472.2170S}], we do not see a reason to apply them in NGC 4958 since we do not have high-quality spectral data along several years to search for variability in the double-peaked component and also because the blue peak is more intense than the red peak, which is expected in relativistic discs due to Doppler boosting effects. We used the Levenberg-Marquardt algorithm with $\xi_i$, $\xi_o$, $i$ and $\sigma$ set as free parameters. We used the same power-law relation for the emissivity of the H$\alpha$ line, assuming q = 3. The resulting values obtained for these parameters are shown in Table \ref{tab:model_parameters}. The disc profile is shown in the bottom panel of Fig. \ref{fig:disc_steps}. 

\begin{table}
\caption{Circular relativistic disc model parameters. We assumed an emissivity law for H$\alpha$ in the form $\epsilon(\xi) \propto \epsilon^{-q}$ with a fixed $q$ = 3.0, based on photoionization model calculations \citep{1989A&A...213...29C, 1990A&A...229..292C, 1990A&A...229..302D, 1990A&A...229..313D}. All errors presented for the parameters below are purely statistical. \label{tab:model_parameters}}

\begin{tabular}{lc}
\hline
Parameter & Fitted value \\
\hline
Inner radius $\xi_{\rm i}$ ($R_{\rm grav}$) & 570$\pm$83 \\
Outer radius $\xi_{\rm o}$ ($R_{\rm grav}$)& 860$\pm$170 \\
Inclination $i$ (degrees) & 27.2 $\pm$0.7 \\
Local broadening $\sigma$ (km s$^{-1}$) & 1310$\pm$70 \\
\hline
\end{tabular}
\end{table}

Although setting the parameter $q$ = 3 is in accordance with photoionization models, we are assuming that H$\alpha$ comes from a region of the accretion disc where this line emission is not saturated (see \citealt{1989A&A...213...29C, 1990A&A...229..313D}). If this is not the case for NGC 4958, the value for $q$ may be slightly different from 3 (indeed, \citealt{2003AJ....126.1720S} found that for a sample of 116 double-peaked galaxies, the average $q$ $<$ 3). So we decided to fit again the double-peaked component with the circular relativistic disc model but leaving $q$ also as a free parameter. However, the resulting $q$ is very uncertain (3$\pm$7) and it does not change by much the results found for the other parameters. Moreover, a visual inspection of the bottom panel of Fig. \ref{fig:disc_steps} shows that the disc profile with a fixed value for $q$ is adequate. Thus, we decided to maintain the results found for the model parameters when fixing $q$ = 3. 

In Fig. \ref{fig:nuc_spectrum_lines}, we show the whole emission in the red portion of the starlight subtracted nuclear spectrum of NGC 4958 together with the fully fitted profile, which is given by the sum of the emissions from the NLR, the BLR and the relativistic disc. We can see that this full fitted profile is in accordance with the observed line emissions. Moreover, the fitted H$\beta$ profile (narrow component plus a central broad component) is adequate for the observed emission, which suggests that no disc emission is seen in this line. 

\section{Discussion} \label{sec:discussion}

This is the first time that a double-peaked H$\alpha$ component is reported for NGC 4958. However, it is not possible to propose that this feature appeared around the year of 2015 (time of our observations) or if it is seen due to the fact that the GMOS-IFU data have a high spatial resolution and less contamination from starlight. For instance, \citet{1989A&A...226...23B} proposed that NGC 4958 had an extremely low-level emission in the galactic nucleus at that time, using observations taken at the ESO 1.52 telescope in La Silla with a slit size of 5 $\times$ 13 arcsec$^2$. Also, the 6dFGS data \citep{2009MNRAS.399..683J}, which was obtained with a fibre aperture of 6.7 arcsec, shows no sign of a broad feature in H$\alpha$. To test if the spatial resolution is important in the case of NGC 4958, we summed all spectra from the GMOS-IFU data cube (FOV = 3.5$\times$5.0 arcsec$^2$) in order to check if we still are able to see the double-peaked component. In fact, this feature disappeared when all the spectra contained in the FOV of the data cube were added together. A similar procedure was performed in NGC 4450 \citep{2000ApJ...541..120H} and NGC 4202 \citep{2000ApJ...534L..27S}: ground-based spectra for both objects did not reveal any sign of a double-peaked component, but this feature was clearly revealed in these galaxies when the authors used HST data. \citet{2003AJ....126.1720S} also claimed that they would have detected more than 116 double-peaked emitters in their sample if better spatial resolutions were available. Thus, higher spatial resolution data is necessary in order to study a possible variation that this component of NGC 4958 may have. 

It is also useful to check how the double-peaked characteristics of NGC 4958 compare to results of other samples. The FWHM is similar to the averaged values taken from the sample of double-peaked emitters of \citet{2003ApJ...599..886E} and is also in accordance with the results presented by \citet{2003AJ....126.1720S}. The same issue applies to the results found for the redshifts of this feature. The peak positions also seem to be in consonance with the results shown in \citet{2003AJ....126.1720S}. Both papers propose that a fraction of the double-peaked sample may be fitted by a relativistic circular disc model, which seems to be the case here for NGC 4958. 

When it comes to the disc parameters obtained with the model, the inclination, the broadening parameter and the inner radius found for NGC 4958 are in accordance with the results presented by \citet{2003AJ....126.1720S} and \citet{2003ApJ...599..886E}. In fact, the inclination is an important parameter to detect the double-peaked profiles. With low disc inclinations, both red and blue peaks are strongly blended \citep{2003ApJ...599..886E, 2017ApJ...835..236S}, while in high inclinations the disc emission may be hidden behind the molecular torus in the context of the unified model \citep{2003ApJ...599..886E}. However, the value found for the outer radius of the disc of this object seems to be small when compared to the \citet{2003AJ....126.1720S} sample, although it is similar to the outer radii results for Arp 102B, Pictor A and 3C 17 \citep{2003ApJ...599..886E}. 

According to the line ratios from the NLR of NGC 4958, this galaxy contains a LINER-like nucleus. This seems to be the case in the majority of the double-peaked emitters \citep{2003ApJ...599..886E}. \citet{2003AJ....126.1720S} claim that line ratios and equivalent widths containing low ionization lines are higher in double-peaked emitters when compared to their parent AGN sample. Moreover, LINERs seem to have a hot ion torus that provides the ``missing'' photons that are needed to explain the disc line emission \citep{1989ApJ...339..742C, 2003ApJ...599..886E, 2008ARA&A..46..475H}. Spectral energy distribution (SED) models show that this is the case for NGC 1097 \citep{2006ApJ...643..652N}. 

In addition to the double-peaked feature, some galaxies also need a central broad component in order to fit the profile of the H$\alpha$ line \citep{2003AJ....126.1720S, 2000ApJ...534L..27S, 2017ApJ...835..236S, 2017MNRAS.472.2170S}. This is clearly the case of NGC 4958. Moreover, this component is also necessary to fit the H$\beta$ emission of this galaxy. The observed H$\alpha$/H$\beta$ ratio for this component is 11$\pm$2. If we assume that the BLR has the same extinction as the NLR, the value of this ratio would be reduced to 7. Both values are in accordance with what was predicted by \citet{2004ApJ...606..749K} for sources with low ionization parameter, which seems to be the case for NGC 4958.

\section{Conclusions} \label{sec:conclusions}

In this paper, we report the detection of a double-peaked profile in the nuclear H$\alpha$ of the galaxy NGC 4958. This is the first time that such a feature is detected in this object. Although emission lines have been mentioned before for this galaxy \citep{1989A&A...226...23B, 2009MNRAS.399..683J}, the use of a high-spatial resolution data with an indubitable quality was probably very important in order to unveil such a component. Below we enunciate our main findings:

\begin{itemize}

\item We found that the nucleus of NGC 4958 has a type 1 LINER-like spectrum. 
\item We detected a double-peaked broad H$\alpha$ emission in the nuclear spectrum. 
\item We propose that the double-peaked broad H$\alpha$ line is the emission from a circular relativistic Keplerian disc with an inner radius $\xi_{\rm i}$ = 570$\pm$83, an outer radius $\xi_{\rm o}$ = 860$\pm$170 (both in units of $GM_{\rm SMBH}/c^2$), an inclination to the line of sight $i$ = 27.2$\degree \pm$0.7$\degree$ and a local broadening parameter $\sigma$ = 1310$\pm$70 km s$^{-1}$. 
\item In addition to the double-peaked broad H$\alpha$ emission, another broad component is necessary to fit the H$\alpha$ profile. This feature is also seen in the H$\beta$ line. We interpret this as a typical BLR emission, which is located beyond the relativistic disc, given that the FWHM of this component is narrower than the FWHM measured for the double-peaked component. The H$\alpha$/H$\beta$ ratio is also in accordance with photoionization calculations for BLR. 

\end{itemize}

\section*{Acknowledgements}

Based on observations obtained at the Gemini Observatory, acquired through the Gemini Observatory Archive and processed using the Gemini IRAF package, which is operated by the Association of Universities for Research in Astronomy, Inc., under a cooperative agreement with the NSF on behalf of the Gemini partnership: the National Science Foundation (United States), National Research Council (Canada), CONICYT (Chile), Ministerio de Ciencia, Tecnolog\'{i}a e Innovaci\'{o}n Productiva (Argentina), Minist\'{e}rio da Ci\^{e}ncia, Tecnologia e Inova\c{c}\~{a}o (Brazil), and Korea Astronomy and Space Science Institute (Republic of Korea). We have also made use of the NASA/IPAC Extragalactic Database (NED), which is operated by the Jet Propulsion Laboratory, California Institute of Technology, under contract with the National Aeronautics and Space Administration. This publication also makes use of data products from the Two Micron All Sky Survey, which is a joint project of the University of Massachusetts and the Infrared Processing and Analysis Center/California Institute of Technology, funded by the National Aeronautics and Space Administration and the National Science Foundation. This work has also made use of the computing facilities of the Laboratory of Astroinformatics (IAG/USP, NAT/Unicsul), whose purchase was made possible by the Brazilian agency FAPESP (grant 2009/54006-4) and the INCT-A.

T.V.R. acknowledges CNPq for financial support under the grant 304321/2016-8. J.E.S acknowledges FAPESP for financial support under the grant 2011/51680-6. We also thank Roberto Menezes and Juliana Motter for a careful review of the manuscript. Finally, we thank the anonymous referee for his/hers comments.

\bibliographystyle{mn2e}
\bibliography{bibliografia}

\appendix

\section{Spectral synthesis results} \label{sec:spectral_synthesis_results}

\begin{figure}

\includegraphics[scale=0.40]{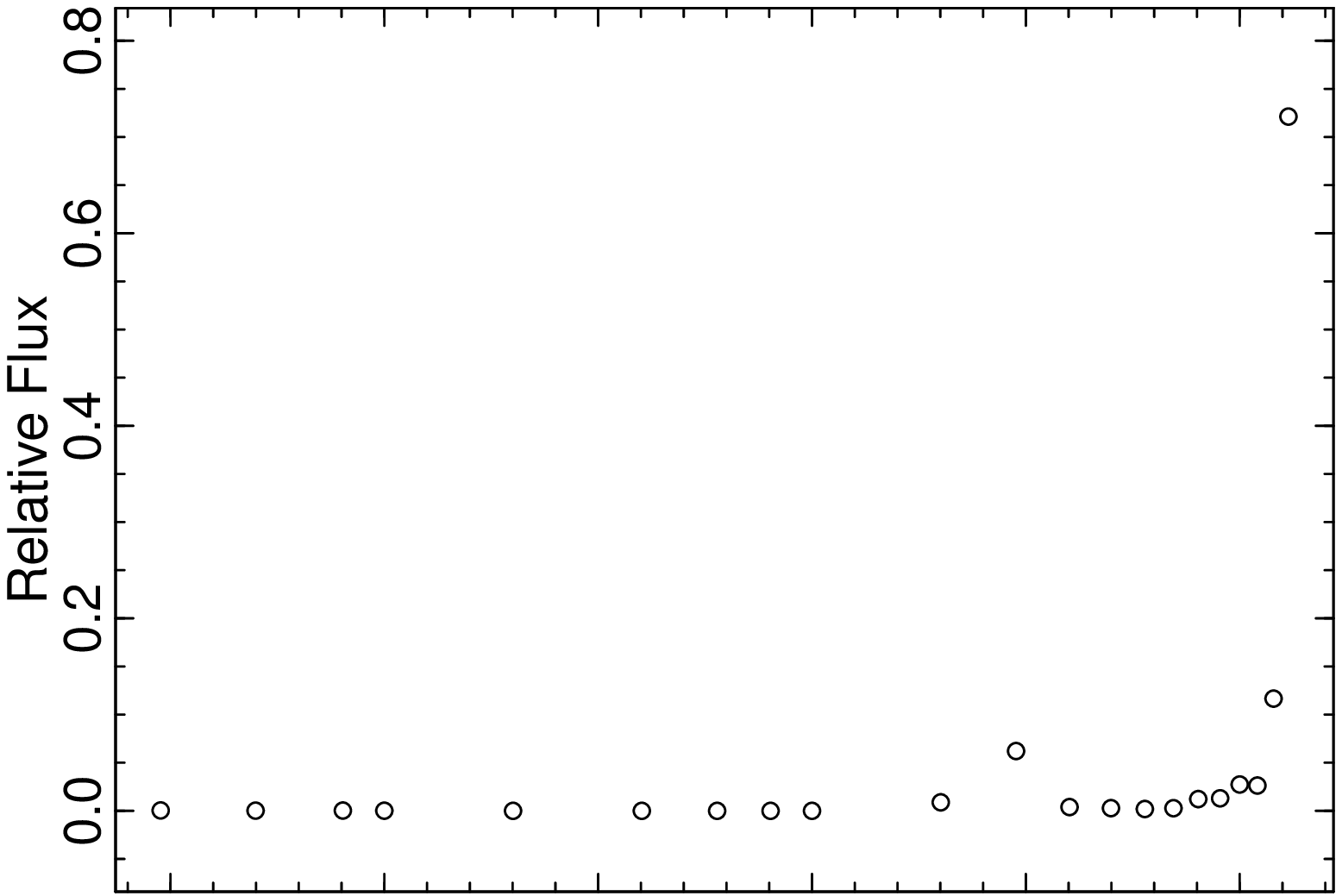} 
\includegraphics[scale=0.40]{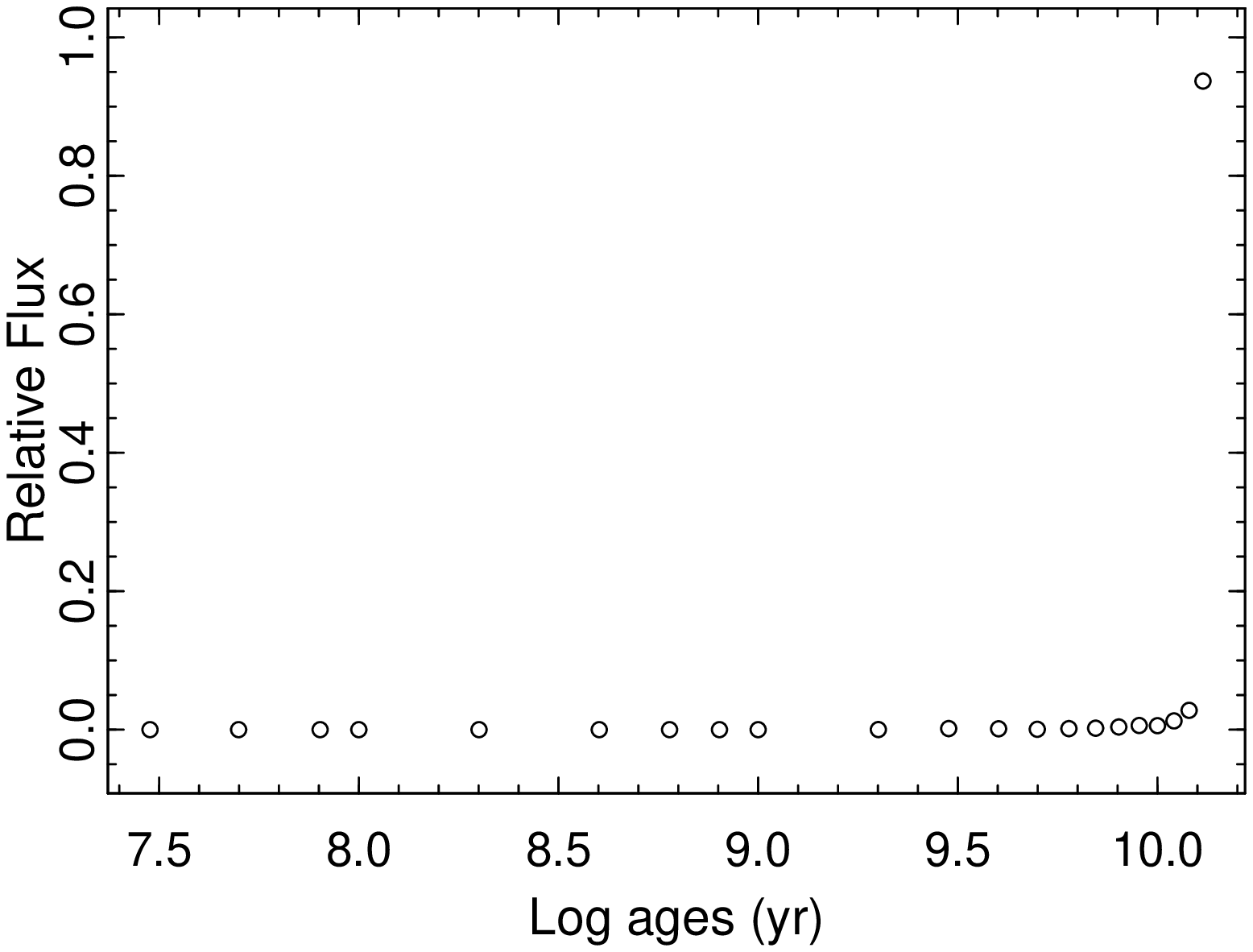} 

\caption{Light fraction of the SSPs as a function of the ages in the nuclear (top) and the circumnuclear (bottom) regions.  \label{fig:ages_pop_stars}}
\end{figure}

\begin{figure}

\includegraphics[scale=0.40]{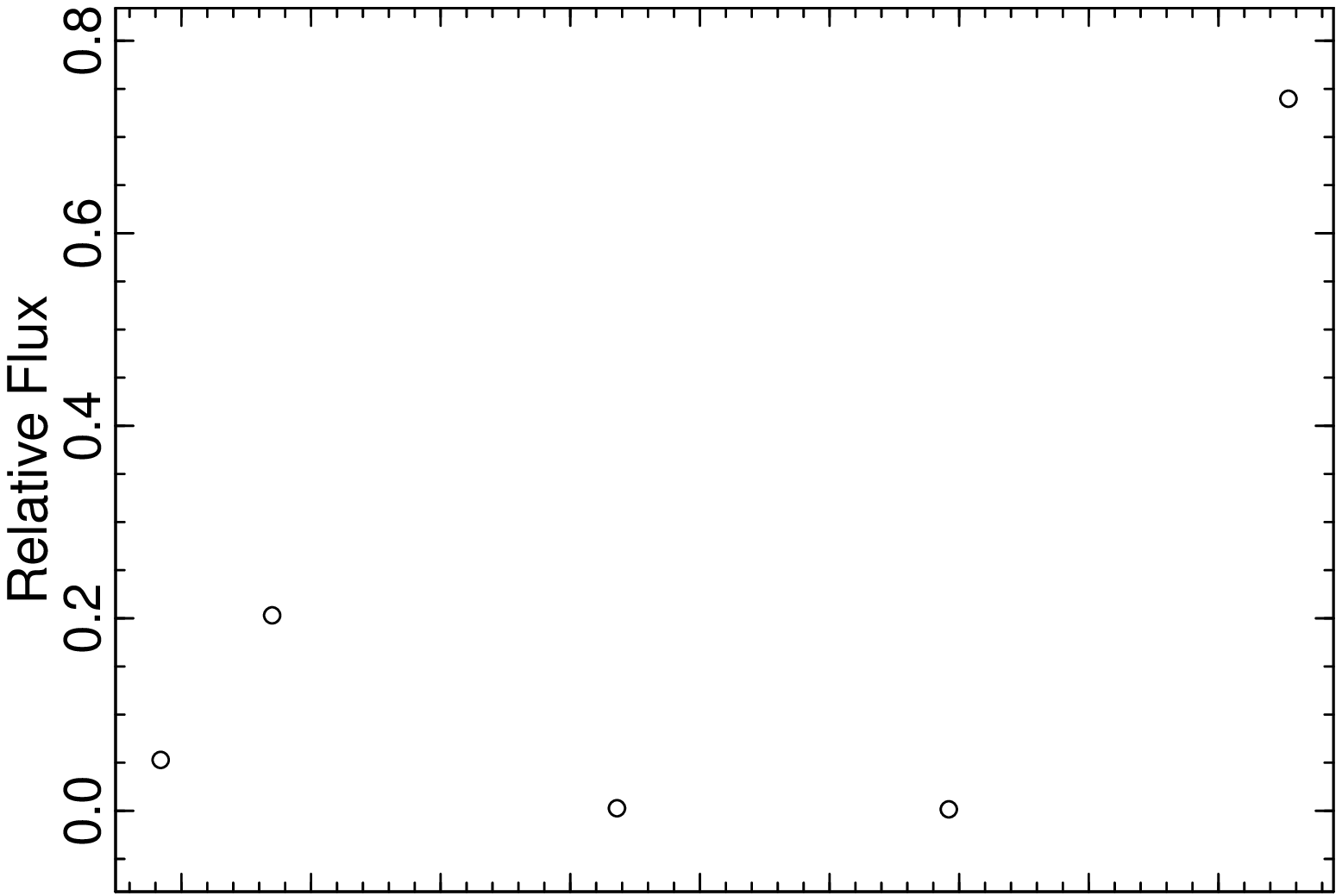} 
\includegraphics[scale=0.40]{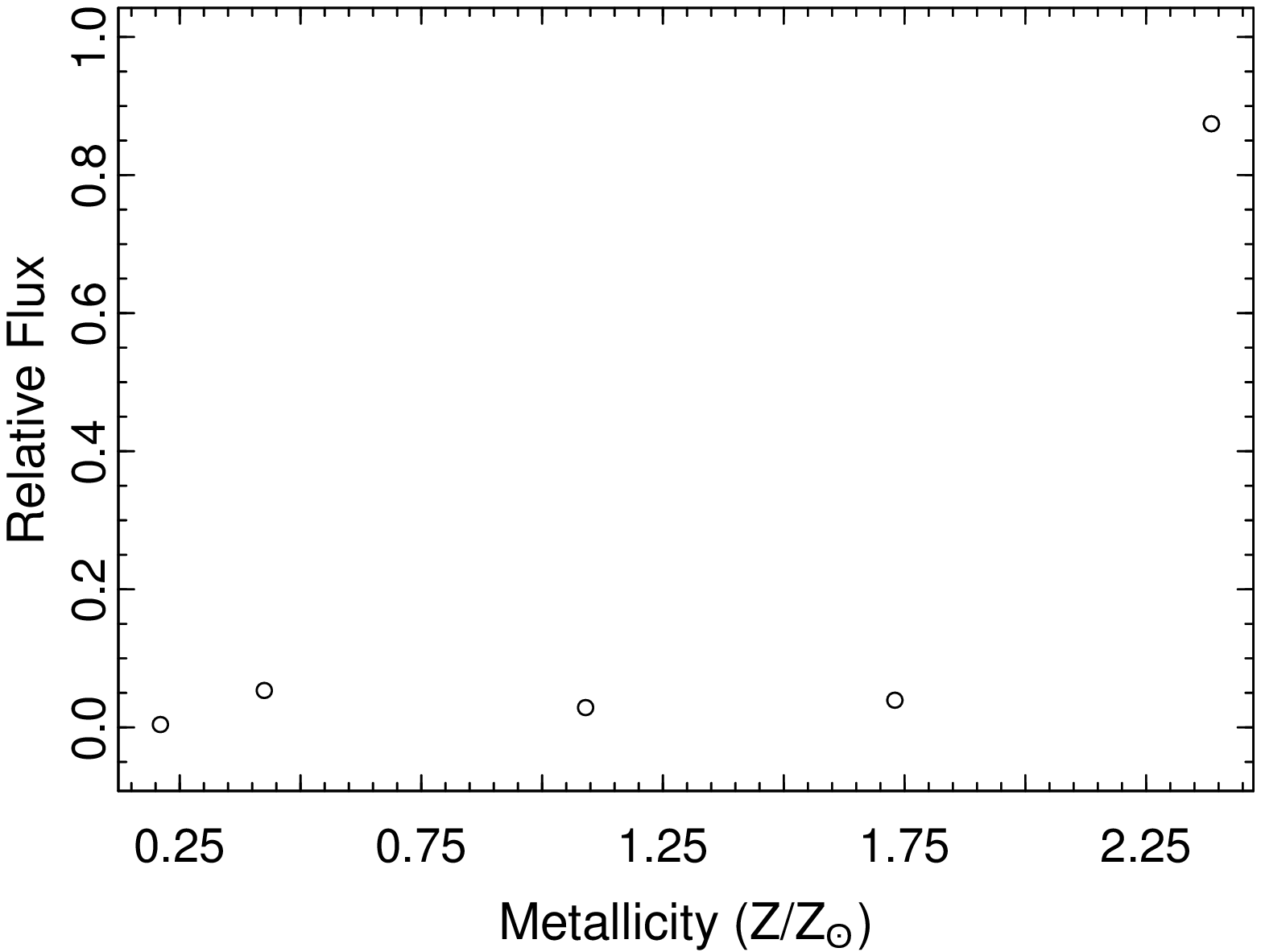} 

\caption{Light fraction of the SSPs as a function of the metallicities in the nuclear (top) and the circumnuclear (bottom) regions. \label{fig:metal_pop_stars}}
\end{figure}

We show in Figs. \ref{fig:ages_pop_stars} and \ref{fig:metal_pop_stars} the results related to the SSPs that best fitted the stellar spectra of the data cube of NGC 4958. The focuses here are the nuclear (top panels) and the circumnuclear (bottom panels) regions. We added all the light contributions of each SSP across the spaxels that are representative of both positions. The relative light fluxes are presented in the figure as a function of the ages and metallicities of the SSPs. 

Most of the starlight comes from stellar populations with old ages ($>$ 10 Gyr) and high metallicity (Z $\sim$ 2.4 Z$_\odot$) in both regions. This is expected for a lenticular galaxy. In addition, for the nuclear region, 25 \% of the starlight is related to SSPs with a low metallicity (Z $<$ 1 Z$_\odot$).  

\end{document}